# Voting Contagion

Dan Braha[1,2] and Marcus A.M. de Aguiar[1,3]

**1** New England Complex Systems Institute, Cambridge, MA, United States,
**2** University of Massachusetts Dartmouth, Dartmouth, MA, United States,
**3** Universidade Estadual de Campinas, Campinas, SP, Brazil

Social influence plays an important role in human behavior and decisions. The sources of influence can be generally divided into external, which are independent of social context, or as originating from peers, such as family and friends. An important question is how to disentangle the social contagion by peers from external influences. While a variety of experimental and observational studies provided insight into this problem, identifying the extent of social contagion based on large-scale observational data with an unknown network structure remains largely unexplored. By bridging the gap between the large-scale complex systems perspective of collective human dynamics and the detailed approach of the social sciences, we present a parsimonious model of social influence, and apply it to a central topic in political science—elections and voting behavior. We provide an analytical expression of the county vote-share distribution in a two-party system, which is in excellent agreement with 92 years of observed U.S. presidential election data. Analyzing the social influence topography over this period reveals an abrupt transition in the patterns of social contagion—from low to high levels of social contagion. The results from our analysis reveal robust differences among regions of the United States in terms of their social influence index. In particular, we identify two regions of 'hot' and 'cold' spots of social influence, each comprising states that are geographically close. These results suggest that social contagion effects are becoming more instrumental in shaping large scale collective political behavior, which is at the core of democratic societies.

# Introduction

The understanding of collective human dynamics in real-life social systems gained increasing attention in recent decades (1-3). At the core of these efforts are models that incorporate a collection of interconnected individuals that change their behavior based on micro-level processes of social influence exerted by their neighbors, but also based on individuals' personal influences independent of social context. The macro-level characteristics of the system emerge as a product of the collective dynamics of these personal influences and micro-level social influence processes. The question of how to separate and measure the effect of social influence is therefore a major challenge for understanding collective human behavior. Although a variety of experimental (4-8) and observational (9-12) studies attempted to address this challenge, identifying the extent of social influence based on large-scale, macro-level observational data in the presence of unknown network structure remains largely unexplored. To close this gap, we present a simple and universal method for measuring social influence, taking the voter model of statistical physics as our basic dynamical system (1, 13-15). We apply our model to understanding the collective dynamics of voting in US presidential elections—a topic at the core of collective political behavior.

The study of electoral behavior has attracted considerable attention by political scientists. Most studies of voting behavior in the United States and other democracies view vote choices as the result of several interrelated attitudinal and social factors (16). Attitudinal factors that reflect short-term fluctuations in partisan division of the vote include evaluations of the candidates' personal qualities and government performance, and orientations toward issues of public policy. Long-term factors, which persist beyond a particular election, include partisan loyalties (17, 18), ideological orientations (19), and social characteristics such as race, religion, social class, and region (16). Recent studies have also elucidated the role of social networks in spreading voting behavior (20). Voters embedded in social networks of friends, family members, neighbors, and co-workers (8) influence each other in terms of voter turnout (8, 21-23) or support particular candidates (20, 24). Social networks enable bounded rational voters to limit the cost of searching for political information (19) by relying on readily available information of their peers. Other sources of political information that were shown to influence citizen attitudes and voting behavior are the mass media (25-29) and a variety of organized efforts at political persuasion such as campaign persuasion (30).

Thus the picture that emerges from the modern history of social science academic voting research suggests that voters are embedded in interpersonal social networks that can increase the likelihood of voting contagion and behavior change; but are also exposed to what we might call external influences affecting their vote decision independent of peer influence. As mentioned above, these external (non-network) influences include various attitudes and orientations, party identification, ideology, campaign persuasion, and exposure to the mass media, such as television and newspapers. The external influences are often skewed in favor of one candidate over another (20). Collectively, the voter's electoral decision can be explained in terms of peer effects and partisan biases conveyed by competing external influences.

A pertinent question here is how to disentangle the effect of social contagion from that of exposure to external influences. This identification problem goes beyond voting. People hold opinions on a multitude of topics that inform alternative courses of action, from crime participation (31) and smoking (32) to riots and protests (33) and financial crises (34). These opinions can be either the result of individual considerations or, when confronted with information that is difficult to acquire or process, influenced by the views of others through social interactions. In this paper we describe a general methodology for detecting behavioral contagion from large-scale observational data. To this end, we take the voter model (1, 13-15) as our basic dynamical system, and extend it to take into account the dynamic response of social networks to external influences. Our model focuses on two characterizations of voting behavior. The first is that of most studies of voting behavior, which consider vote choices to be driven, as outlined above, by various attitudinal factors and other external pressures. The second—from complex systems science and recent observational and face-to-face studies—is that of internal self-reinforcing dynamics where voters' opinions are changed under the influence of their peers. Incorporating both, we construct a universal representation of the largest scale system behavior when there is both external and interpersonal influence. The extended voter model is able to reproduce remarkably well statistical features and patterns of almost a century of county-level U.S. presidential election data. More importantly, our model presents a general framework for detecting social contagion from large-scale election return data, and can be applied more generally to many different systems that involve social contagion.

Here, electoral votes cast for U.S. presidential candidates at the county level are analyzed, covering the period of 1920 through 2012. Counties are grouped by state, and the corresponding distributions of the fraction of votes (vote-share) in a county for the Democratic candidate in an election year are studied. Fig. 1 shows the county vote-share distributions for various states and election years. The data indicates that there is a wide variation in the characteristics of voting behavior with no apparent pattern of voting dynamics in time or geographical space. Here we show that much of this observed variability of county vote-shares may be best explained by fluctuating peer influences across time and space. Although the study of collective opinion dynamics, including voting behavior, has been the focus of discussion in the context of modelling and identification of universal patterns of behavior (35-45), the mechanisms leading to the diverse spatiotemporal variation in voting patterns shown in Fig. 1 are poorly understood. The model presented below provides a parsimonious quantitative framework that is capable to explain and reproduce the full range of empirical county vote-share distributions for all states and election years. Using the model, we develop an index of social influence that enables us to examine and reveal remarkably robust patterns of spatial and temporal variation in social influence over 92 years of US presidential elections.

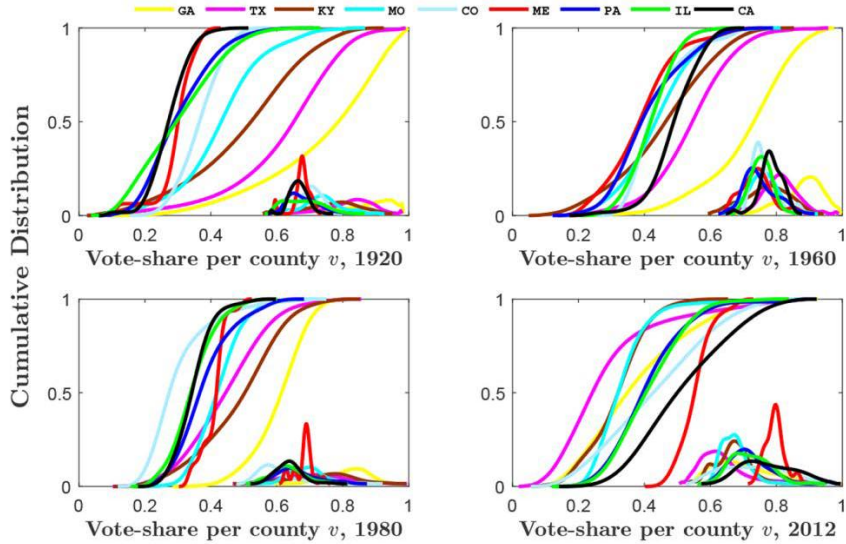

**Fig. 1. Observed county vote-share distributions, 1920-2012.** The observed distributions of Democratic vote-share per county are presented for states with the greatest number of counties, for each of the nine census divisions established by the U.S. Census Bureau. Here plots are shown for various presidential elections from 1920 to 2012. The vote-share per county is measured as the percentage of the vote in the county received by the Democratic candidate. The figure shows the plot of the cumulative distributions. The insets show the corresponding probability density functions (for clarity the x-scale has been shifted to right and contracted). Curves are based on kernel estimation with Gaussian kernels.

## Results

### Model of voting contagion under external influences

To model the dynamics of elections, we consider a network with $N$ nodes representing voters, and links between pairs of nodes representing interpersonal influence among voters. To represent external influences, we add nodes that influence others, but are not themselves influenced, i.e. "fixed" nodes. The number of fixed nodes that are biased in favor of the first candidate (named '0') is $N^0$ and the number biased in favor of the second candidate (named '1') is $N^1$. The effective strength of the external influences biased towards the two candidates is given by the number of these nodes. Thus, we consider a network with $N + N^0 + N^1$ nodes. Each node has an internal state which can take only the values 0 and 1, representing whether the voter chooses the first or second candidate; or, for fixed nodes, whether the node is biased in favor of the first or second candidate. We assume that the $N$ *free* nodes (voters) change their internal state according to a dynamical rule: At each time step a random free node is selected and its state is updated with probability $1 - p$ by copying the state of one of its connected neighbors, chosen at random from all nodes; and with probability $p$ the state remains the same. The $N^0$ and $N^1$ nodes that are biased towards the first and second candidate remain fixed in state 0 and state 1, respectively. Copying the state of a connected free node represents mutual influence among friends, neighbors, and family members; while copying from a fixed node represents the influence of external factors such as exposure to mass media, candidate qualities, issue orientation, the incumbent party's performance in office; or deep-rooted values that define and shape an individual's party identification and ideology. Analytically extending $N^0$ and $N^1$ to non-integer values enables modeling arbitrary relative strength of external to internal influence (see Materials and Methods). The number of fixed nodes can be interpreted in several ways: Fixed nodes can be viewed as autonomous voters who do not observe their neighbors' vote choices. Accordingly, the number of fixed nodes may reflect the proportion of the electorate with very weak or no ties to other members of a social network; or as the proportion of the electorate who are "opinion leaders" who interpret media messages and pass them on to "opinion followers" (17). The number of fixed nodes can also be seen as a device for representing the external forces that inhibit the effect of peer influence, such as party identification, ideology or any other source of political information— this is the view that we shall adopt here. We note that in this model external influences of opposite partisan biases do not cancel; instead larger $N^0$ and $N^1$ reflect increasing probability that external influences determine the choices of voters independent of the voting behaviors of others. The model assumes that there are many external sources of competing political information, and that over the election period in question the sources are persistent in their proportion of partisan biases regarding the two-major party candidates though vary in the way they influence individual voters' choices. Election years that are consistently biased towards the first (second) party's candidate would be represented by $N^0$ greater (smaller) than $N^1$. We have previously proposed this model as a widely applicable theory of collective behavior of complex systems (14-15, 46-48). Here we extend the model to divulge the phenomenology of voting contagion in electoral voting behavior. Successful matching to election data will also be a confirmation of the universality of this theory.

The behavioral model can be solved exactly for a fully connected network. At equilibrium, the probability of finding the network in the global state of $k$ voters in state 1 (i.e. voting for candidate 1) is given, independently of the initial state, as follows (see derivation in Materials and Methods):

$$\rho(k) = \frac{\binom{N^1+k-1}{k}\binom{N+N^0-k-1}{N-k}}{\binom{N+N^0+N^1-1}{N}}. \qquad [1]$$

where $N$ is the number of voters, $k$ is the number of voters is state 1 and $\binom{n}{k}$ are binomial coefficients. Analytically extending $N^0$ and $N^1$ to non-integer values enables modeling arbitrary relative strength of external to internal influence (see Materials and Methods). In this case, the solution in Eq. 1 remains the same, with the difference that factorials must be replaced by gamma functions. Indeed, as we move around in the $(N^0, N^1)$-parameter space, the stationary distribution in Eq. 1 exhibits strikingly different shapes depending on the strength of the external forces, $N^0$ and $N^1$, compared to the strength of interactions within the network, and the relative bias of the external influence toward the first or second candidate—from skewed unimodal distributions with intermediate peaks to peaks at all nodes 1 or all nodes 0 to bimodal and uniform distributions (see Fig. 6 in Materials and Methods). Interestingly, Eq. 1 remains valid for other network topologies if $N^0$ and $N^1$ are re-scaled according to the degree distribution (see Materials and Methods).

In this paper we are mostly interested in the fraction of voters (vote-share) that voted for a candidate rather than the actual number of voters. Thus, we define the vote-share for candidate 1 as the scaled variable $v = k/N$. The mean and variance of $v$ can be computed from Eq. 1 as follows

$$\mu_v = \frac{N^1}{N^0 + N^1} \qquad [2]$$

$$\sigma_v^2 = \overbrace{\frac{\mu_v(1-\mu_v)}{N}}^{\text{External influence}} \overbrace{\left(\frac{N}{N^0+N^1+1} + \frac{N^0+N^1}{N^0+N^1+1}\right)}^{\text{Social Influence}} \qquad [3]$$

The variance of vote-shares in Eq. 3 has an appealing interpretation. When peer influences are very weak compared to external forces $(N^0, N^1 \to \infty)$, the variance of vote-shares becomes $\sigma_v^2 = \mu_v(1-\mu_v)/N$. This is the variance of vote-shares that one would expect if all voters make independent decisions, each with probability $\mu_v$. The second term on the right side of Eq. 3 is a decreasing nonlinear function of the external influence parameters, and thus provides us with a way of detecting and isolating the effect of social contagion. This index can be applied to understanding how social influence changes across states and over election years.

The unknown external influence parameters $N^0$ and $N^1$ for any state in any election year can be estimated using the sample vote-shares across counties as follows. Suppose a particular state has $n$ counties, and let $v_i$ be the fraction of voters in the $i^{\text{th}}$ county that voted for candidate 1, and $N_i$ be the total number of votes cast for all candidates in the county. We assume that all counties are influenced by the same external parameters and that the distribution is in equilibrium. Using these assumptions, we view each county as a subnetwork with $N_i$ nodes and external parameters $N^0$ and $N^1$. Using Eq. 2, the mean of vote-shares in county $i$ does not depend on $i$, and is equal to $\mu_i = \mu = N^1/(N^0 + N^1)$. We thus estimate $\mu_i$ by simply taking the sample average $\hat{\mu}$ of vote-shares across all $n$ counties. For the variance of vote-shares $\sigma_i^2$ in county $i$, a crude estimate based on the single observed vote-share data point $v_i$ is provided by $(v_i - \hat{\mu})^2$. Obviously, this estimate is imperfect and we define the residual between $\sigma_i^2$ and the estimate of $\sigma_i^2$

$$\varepsilon_i = (v_i - \hat{\mu})^2 - \sigma_i^2 \qquad [4]$$

Using Eqs. 3 and 4, we define a system of nonlinear estimation equations that relates $(v_i - \hat{\mu})^2$, the estimate of $\sigma_i^2$, to the external influence parameters $N^0$ and $N^1$:

$$(v_i - \hat{\mu})^2 = \frac{\mu(1-\mu)}{N_i}\left(\frac{N_i}{N^0 + N^1 + 1} + \frac{N^0 + N^1}{N^0 + N^1 + 1}\right) + \epsilon_i \quad i = 1, \cdots, n \qquad [5]$$

The estimation procedure first estimates $\mu$ on the right hand side of Eq. 5 by $\hat{\mu}$, and then select the sum of parameters $N^0 + N^1$ that minimizes $\sum_{i=1}^n \varepsilon_i^2$ in Eq. 5. The least squares estimate is given by

$$\widehat{N}^0 + \widehat{N}^1 = \frac{n\hat{\mu}(1-\hat{\mu}) + \sum_i \frac{(v_i - \hat{\mu})^2}{N_i} - \hat{\mu}(1-\hat{\mu})\sum_i \frac{1}{N_i} - \sum_i(v_i - \hat{\mu})^2}{-\hat{\mu}(1-\hat{\mu})\sum_i \frac{1}{N_i} + \sum_i(v_i - \hat{\mu})^2 + \hat{\mu}(1-\hat{\mu})\sum_i \frac{1}{N_i^2} - \sum_i \frac{(v_i - \hat{\mu})^2}{N_i}} \qquad [6]$$

Eq. 6 and the condition $\hat{\mu} = \widehat{N}^1/(\widehat{N}^0 + \widehat{N}^1)$ fully determine the estimated external influence parameters. We can use Eq. 3 and the estimate in Eq. 6 to obtain an index of social influence for each state in each election year:

$$\frac{\bar{N}}{\widehat{N}^0 + \widehat{N}^1 + 1} + \frac{\widehat{N}^0 + \widehat{N}^1}{\widehat{N}^0 + \widehat{N}^1 + 1} \qquad [7]$$

where $\bar{N} = \sum_i N_i/n$ is the average number of voters per county.

For the U.S. presidential elections from 1920 to 2012, we find that $\widehat{N}^0, \widehat{N}^1 \gg 1$ for all states and across election years. Driven by this fact and the fact that the total number of voters $N_i$ in any county $i$ in any given election year is large, we obtain in the limit $N_i \to \infty$ that the stationary distribution in Eq. 1 is approximately a Gaussian distribution (see Materials and Methods). More specifically, we approximate the asymptotic vote-share distribution in county $i$ by a Gaussian $\rho(v_i) = 1/\sqrt{2\pi\sigma_i^2}\, e^{[-(v_i-\mu_i)^2/2\sigma_i^2]}$ with mean $\mu_i = \mu = N^1/(N^0+N^1)$ and variance $\sigma_i^2 = \mu(1-\mu)\left(\frac{1}{N^0+N^1} + \frac{1}{N_i}\right)$. In our case we observe the vote-shares of $n$ different counties of a state, and are interested in the probability distribution of observations across all counties (see Fig. 1). A plausible model for this distribution of vote-shares $v$ per county is to describe it as a Gaussian scale mixture (49) with $n$ different components (representing counties), each distributed as a normal distribution with the same mean $\mu$ and variance $\sigma_i^2$, as specified above. This mixture is unimodal with a mode at $\mu$, skewness value $\beta_1 = 0$, kutosis value $\beta_2 = 3 \sum_{i=1}^n \frac{1}{n}\sigma_i^4 / \left(\sum_{i=1}^n \frac{1}{n}\sigma_i^2\right)^2$, and variance $\sigma_v^2 = \sum_{i=1}^n \frac{1}{n}\sigma_i^2$. Using the Pearson system, the Gaussian scale mixture can be approximated by a Student's $t$-distribution (50). More specifically, let $v$—the vote-share per county in the state—be a Gaussian mixture of $n$ components with equal mean $\mu$. Let $c_0 = 2\sigma_v^2\beta_2/(5\beta_2-9)$ and $c_2 = (\beta_2-3)/(5\beta_2-9)$ be the Pearson coefficients, $\alpha = \sqrt{(1-c_2)/c_0}$, and $m = (1-c_2)/c_2$. Then, the scaled and shifted mixture $\alpha(v-\mu)$ is approximately distributed as a Student's $t$-distribution with $m$ degrees of freedom (50). Notice that $\mu, \sigma_i^2, \sigma_v^2, c_0, c_2, \alpha$ and $v$ are functions of the external influence parameters $N^0, N^1$, the number of counties $n$ in the state, and the total number of votes $N_i$ in each county, and thus can be estimated from them. A main result of our paper is that for all states and across election years the number of degrees of freedom $m \gg 100$, and thus the scaled and shifted vote-share $\alpha(v-\mu)$ is predicted to approach the standard normal distribution.

**Empirical results of US presidential elections: 1920 to 2012**

Our analysis is based on US presidential election data from 1920 to 2012 (51). States with less than 10 counties (i.e., Connecticut, Delaware, Hawaii, and Rhode Island) and Washington D.C. were excluded from analysis. For each state, in every election year, the data includes information on the number of counties $n$ for which vote-share data was available, the vote-share $v_i$ in county $i$, and the total number of votes cast for all candidates in county $i$, $N_i$. The external influence parameters $N^0$ and $N^1$, and the distribution parameters $\alpha$ and $m$ were estimated for all states and every election year. Using these parameters, we constructed the probability distribution of the scaled and shifted vote-share quantity $\alpha(v_i - \mu)$, and compared it with the predicted normal distribution. Figure 2 shows that this theoretical prediction fits remarkably well for most states and election years, representing almost a century of county-level U.S. presidential election data, and is consistent with observations in other countries (52).

We can use the index of social influence defined in Eq. 7 to examine the level of social interactions across states and election years. In the upper panel of Figure 3 we show the histogram of the social influence index aggregated over all states and election years. The histogram shows a right-skewed log-normal-like distribution. This means that while the bulk of the distribution occurs for small values of social interactions, the electorate in US presidential elections is at times highly volatile and subject to wide swings of social contagion effects higher than the typical value. This is reflected by the highly right-skewed tail of the histogram. Here we find that the log-logistic provides a slightly better fit relative to the log-normal distribution. The log-logistic is a heavy-tailed distribution similar in shape to the log-normal distribution, but with heavier tails (53).

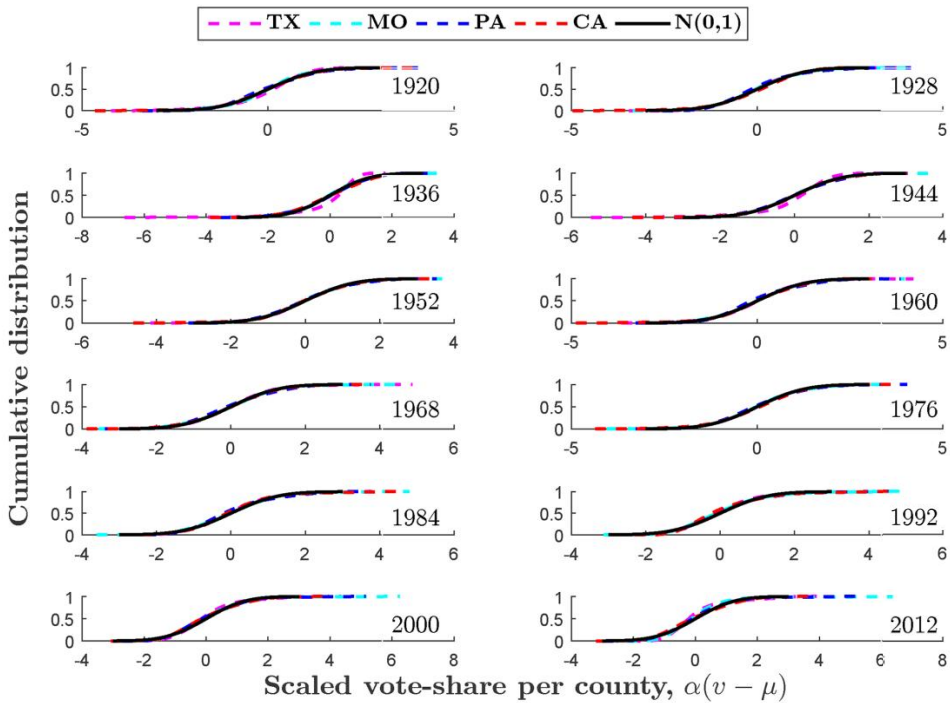

**Fig. 2. Scaled vote-share distributions and predicted curves, 1920-2012.** Without loss of generality, curves are presented for states with the greatest number of counties in each of the four census regions—Northeast,

Midwest, South, and West—of the U.S. The figure shows the plot of the cumulative distributions. Observed values (dashed lines) are based on kernel estimation with Gaussian kernels. Solid lines are Gaussian distributions with mean 0 and variance 1. The scaled vote-shares are calculated from the estimated external influence parameters $N^0$ and $N^1$. The goodness of fit of the Gaussian relative to the empirically observed county vote-share distributions was determined by using a Kolmogorov-Smirnov test. The fit of the model is excellent—the test fails to reject the normality null hypothesis at the 5% significance level, for 95% of all states in all election years; and at the 1% significance level, for 98% of all states in all election years.

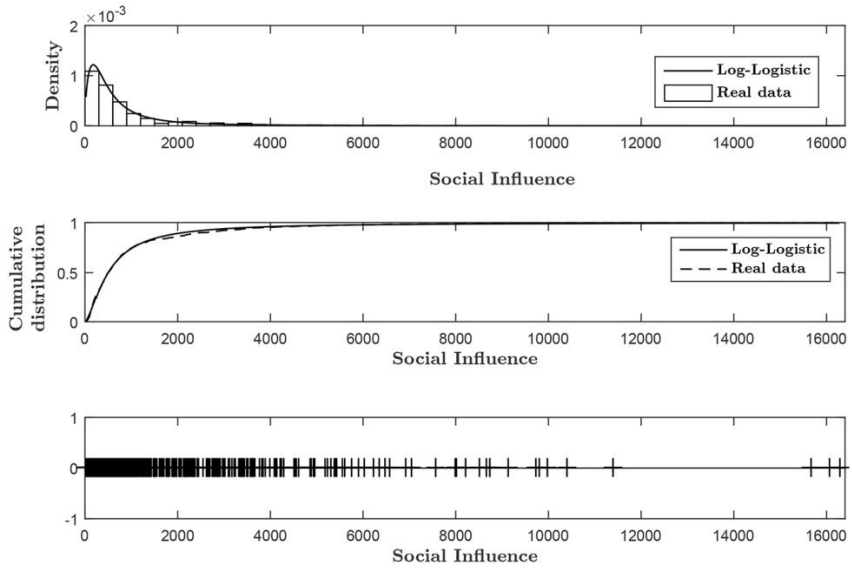

**Fig. 3. Distributions of Social Influence and best-fit curves, 1920-2012.** The upper and middle panels of Figure 3 show the histogram and cumulative distribution of the Social Influence index (using Eq. 7), aggregated over all states and election years. The lower panel shows the spread of the Social Influence index. The broad distributions are best fitted by the log-logistic distribution, often used for analyzing skewed data. The goodness of fit of the log-logistic distribution was determined by using a Kolmogorov-Smirnov test. The fit of the model is very good (p = 0.16).

To analyze the electoral dynamics, we examine the spatial and temporal variation in social influence from 1920 to 2012. First, we examine the evolution of social influence over time. Fig. 4 shows the time series of the average social influence for each of the nine U.S. census divisions (panels a-f) along with the time series of (normalized) social influence averaged over all U.S. states (panel g). To enable the comparison of the various time series, all data are normalized $Z$–scores. Specifically, for each individual time series we express the social influence in terms of standard deviation from their mean, calculated from 1920 to 2012. We use hierarchical clustering to identify clusters of U.S. census divisions with the highest within-cluster time-series correlation and the greatest between-cluster time-series variability. The result from the hierarchical clustering suggests three clusters: two main clusters (arranged in panels d and f) with within-cluster average correlations of 0.824 and 0.909, and a relatively high between-cluster average correlation of 0.775; and a singleton cluster (New England) with a relatively low between-cluster average correlation of 0.1071. Remarkably, we find that despite variations in social influence across states and divisions, the normalized time series pertaining to the overwhelming number of states (with the exception of the three-state region of New England analyzed here) collapse on a very similar curve. Indeed, as can be seen, the normalized curves in panels d and f show a very similar pattern, which is also similar to the observed temporal pattern of social influence when averaged over all states (Fig. 4g). That is, the pattern in Fig. 4g shows a monotonic upward trend, which means that social influence increases through time (Mann-Kendall test, $p \approx 7 \cdot 10^{-6}$). Moreover, the period of 1984–2012 displays much higher levels of social influence when compared with the period of 1920–1980, which displays lower levels of social influence (Mann-Whitney U-test, $p < 1.3 \cdot 10^{-4}$, see Fig. S1). New England is an apparent exception to this pattern (Fig. 4a-b). However, this exception may be explained by the historical events and our model. One of the most unique characteristics that makes New England, as a political region in America, different from other regions is its town meeting form of government—a local institution that did not spread to other states (54). The town meeting is the legislative assembly of a town in which qualified voters make laws in face-to-face communal decision making (54). Town meetings defined New England's politics until the middle decades of the 20$^{th}$ century. This was changed in 1962 with the Supreme Court's "one person, one vote" decisions, which resulted in shifting power dynamic away from most small towns that practiced town meetings, face-to-face interactions, to cities that adopted representative politics (54). Thus the relative high levels of social interaction observed in New England prior to the 1960 election (see Fig. 4b)—contrary to the patterns observed in other regions—correspond to the period in which town meetings—a powerful platform of social influence via face-to-face, communal, decision making—had wide legislative powers. This was followed by a sharp decline in relative social influence (see Fig. 4b) after the Supreme Court's "one person, one vote" decision, which had the effect of shifting the power from face-to-face communication and social interaction to representative politics. This political transition changed not only the relative level of social interactions—and thus the variability of the vote-share distributions—but also impacted the partisan bias—hence the mean of the vote-share distributions—towards the Democrats (54).

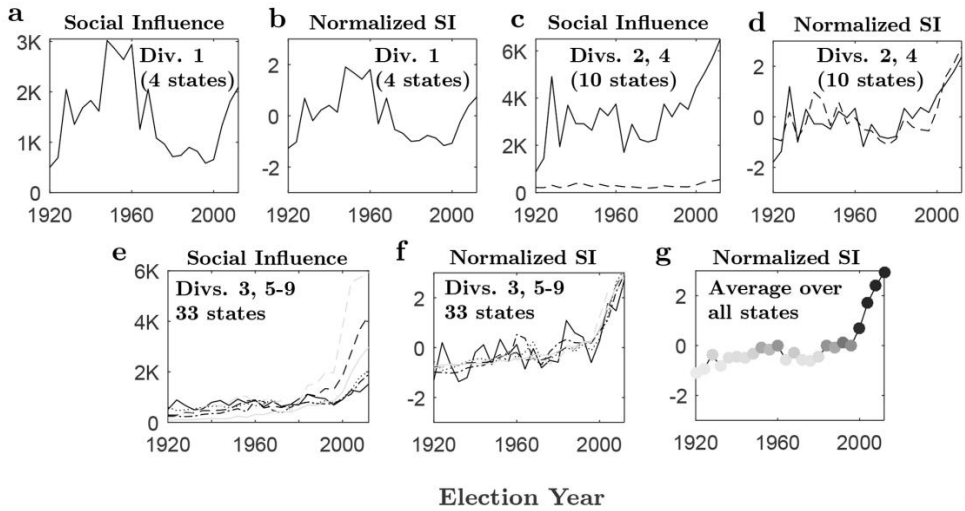

**Fig. 4. Evolution of Social Influence, 1920-2012.** Time series of social influence averaged over states, and their normalized versions (see text), are shown for each of the nine U.S. census divisions: **a-b)** Division 1, New England—Maine, Massachusetts, New Hampshire, and Vermont; **c-d)** Division 2, Middle Atlantic (Solid line): New Jersey, New York and Pennsylvania. Division 4, West North Central (Dashed line): Iowa, Kansas, Minnesota, Missouri, Nebraska, North Dakota and South Dakota. **e-f)** Division 3, East North Central (Solid line): Illinois, Indiana, Michigan, Ohio and Wisconsin. Division 5, South Atlantic (Dashed line): Florida, Georgia, Maryland, North Carolina, South Carolina, Virginia and West Virginia. Division 6, East South Central (dotted line): Alabama, Kentucky, Mississippi and Tennessee. Division 7, West South Central (Dash-dot line): Arkansas, Louisiana, Oklahoma and Texas. Division 8, Mountain (Gray solid line): Arizona, Colorado, Idaho, Montana, Nevada, New Mexico, Utah and Wyoming. Division 9, Pacific: Alaska, California, Oregon and Washington (Gray dashed line). The nine U.S. census divisions are clustered according to the correlation between their respective normalized social influence profiles. The corresponding average pairwise correlations are 0.824 and 0.909 for the clusters in Fig. 4c-d and Fig. 4e-f, respectively. Panel 4g shows the time series of normalized social influence, averaged over all U.S. states. The $Z$–scores are mapped to colors from white ($z = -1.1$, below the mean) to black ($z = 2.9$, above the mean). A clear pattern of high social influence (positive or near-zero $Z$–scores) follows a period (1920–1980) of low social influence (negative $Z$–scores). The 1984 break date—separating low from high levels of social contagion—is identified by the Mann-Whitney U-test, which is applied for different potential breaks within the range 1920-2012 (see Fig. S1).

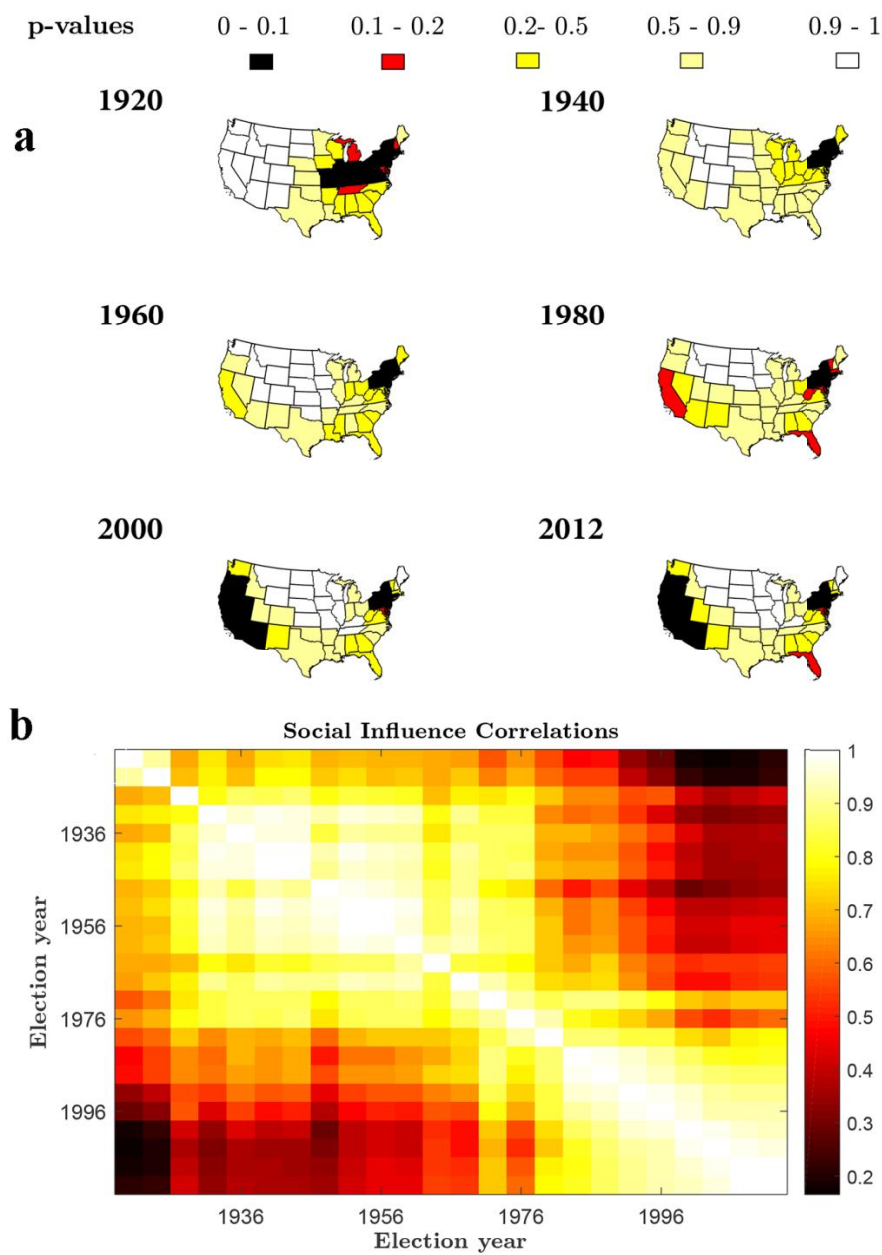

**Fig. 5. Spatial variation of social influence and its change over time, 1920-2012. a)** Hot spot analysis of social influence for sample maps of US presidential elections (see Figs. S2-S25 and Movie S2 for a complete analysis). The colored areas reflect the significance (p-value) of local concentration of social influence for each state. The p-values for each state are derived from a random permutation test of local clustering using the Getis-Ord Local $G_i^*$ statistic. This analysis was performed with a contiguity spatial weight matrix that indicates whether states share a boundary or not. The variable of concern is the social influence index calculated using Eq. 7 in the main text.

Low p-values (p-value≤0.1) indicate statistically significant high levels of social influence at a state and its surrounding neighbors (hot spots). High p-values (p-value≥0.9) indicate statistically significant low levels of social influence at a state and its surrounding neighbors (cold spots). **b)** Heatmap of the correlation coefficients between pairs of election years characterized by their state-level social influence profiles.

As a further support for the usefulness and consistency of our model, we examine how the spatial variation of social influence across states changes over time. We can characterize each election as a vector of state-level indices of social influence (using Eq. 7), and measure the similarity between each pair of elections by the corresponding correlation coefficient (Fig. 5b). This type of analysis, combined with the findings in Fig. 4, reveals intriguing patterns that go beyond short-term fluctuations in partisan division of the vote. Hierarchical clustering of the elections by the social influence correlation distance shows several marked clusters of highly similar election years (see Fig. 4b): 1932–1972, 1984–2012, and three smaller clusters 1920–1924, 1928, and 1976-1980. Remarkably, these clusters of social interactions correspond nicely with the partitioning of American history into distinct party systems (55).

There have been six party system periods in American history, separated by relatively significant change in party loyalties (55-59). Clustering analysis reveals that during 1932-1972, external forces (in the form of attitudes, orientations, party identification, or ideology) are strong compared to social/peer influences, and indicate a stable long-term electorate phase. This result is plausibly supported by the historical account. The stock market crash of 1929 and the ensuing depression signaled the realignment of the fifth party system from a Republican to Democratic majority with the election of 1932 and the New Deal coalition (55, 57). The change was also influenced by demographic changes of rising American electorate of African Americans, blue collar workers, Catholics and urban ethnics, and a shrinking Republican base of white Protestants, small town residents, farmers, and middle class businessmen (57). The distinction between external and social influence stands despite some fluctuations in Republican vs. Democratic selections.

The cluster 1976-1980 identified by our clustering analysis (see also Fig. 5b) suggests that the elections of 1976 and 1980 formed a transition period to the post-New Deal era of weakened partisanship among the voters (57). This transition period corresponds to the Watergate scandal and Nixon's resignation in 1974, and Democrat Jimmy Carter's victory in the 1976 presidential election.

Whereas the previous fifth party system was characterized by strong party loyalties and partisan attachments, the sixth party system (overlapping with the 1984–2012 cluster in Fig. 5b) is characterized by electoral dealignment—the weakening of party loyalties among voters (57, 59, 60), reduced political involvement (61), and the critical role of voters' personal social interaction networks in determining vote choices (20). As partisanship declined and more voters became independents (55), inter-election vote swings increased (57). Moreover, the external influence of television and newspaper declined as the media were considerably less likely to be sources of partisan-biased information (20, 57, 62). This led to a period of strong competition where neither Democrats nor Republicans created a true majority party, resulting in alternating control of the presidency, split-ticket voting, and divided government. These trends seem to be consistent with our model, which shows higher levels of social contagion for the 1984–2012 period (Fig. 4g), relative to the 1920–1980 period, combined with the long-term stability of social influence patterns indicated by the high levels of association between the 1984–2012 elections (Fig. 5b). The high levels of social interactions observed in the 1984–2012 period (Fig. 4g) account for an increasing volatility and variability of the vote-share distributions (via Eq. 3). This is seen in the historical account: The Republicans won the presidency with the victories of Ronald Reagan and George H. W. Bush in 1980, 1984, and 1988, and regained control of the Senate from 1981 to 1987 for the first time in almost 30 years. The Democrats regained control of the presidency with Bill Clinton in 1992 and 1996 whereas the Republicans won control of the Congress from 1994 to 2006 for the first time in 40 years. In 2000, Republican George W. Bush defeated Democratic Al Gore in the closest election in modern U.S. history. Although Bush won reelection in 2004, Democrats won control of Congress in 2006, and Democrat Barack Obama was elected in 2008. Although it would seem that Obama's victories in the 2008 and 2012 suggest a critical realignment of the party system, the Republicans regained control of the House in 2010 by their biggest landslide since 1946, and control of Congress in 2014, with the largest Republican majority in the House since 1928.

Movie S1 in Supplementary Material shows maps (excluding Alaska and Hawaii), color-coded by levels of social influence, for all election years. In order to better characterize the spatial patterns of social influence observed in Movie S1, we apply a variety of spatial statistical data analysis methods. First, we utilized a random permutation test of spatial autocorrelation using the Moran's $I$ statistic (63, 64). The random permutation tests suggest (see Table S1) the presence of significant positive spatial correlation, for all election years, between states' own levels of social influence and the levels of their neighbors as indicated by the level of significance (p-value) shown in the third column of Table S1. This analysis was performed with a contiguity spatial weight matrix (row normalized) that indicates whether states share a boundary or not. While the Moran's $I$ statistic indicates that the spatial distribution of high and/or low values is more spatially clustered than would be expected if underlying processes were random, it does not identify unexpected spatial spikes of high or low social influence values. We thus applied random permutation tests of spatial clustering using the Getis-Ord General $G^*$ statistic (65, 66). The tests indicate (see Table S2) that social influence is significantly concentrated in space as shown by the significance levels (p-value) in the third column of Table S2. That is, for all election years, the observed Getis-Ord General $G^*$ is larger than the expected General $G^*$, indicating that the spatial distribution of high social influence values is more spatially clustered than would be expected if underlying spatial processes were truly random.

In order to identify where high or low values of social influence cluster spatially, we further applied a random permutation test of local clustering using the Getis-Ord Local $G_i^*$ statistic (65, 66). Low p-values of the random permutation test indicate statistically significant high levels of social influence at a state and its surrounding neighbors (hot spots). High p-values indicate statistically significant low levels of social influence at a state and its surrounding neighbors (cold spots). This analysis was performed with a contiguity spatial weight matrix that indicates whether states share a boundary or not. The corresponding maps of hot spot analysis, for all election years, are presented in Figs. S2-S25 and Movie S2. A sample of these maps of social influence clusters is presented in Fig. 5a. The colored areas in Fig. 5a reflect the significance (p-value) of local concentration of social influence for each state, derived from the random permutation test of local clustering using the Getis-Ord Local $G_i^*$ statistic. The maps shown in Fig. 5a (see Figs. S2-S25 and Movie S2 for a complete analysis) enable to identify unusual geographical concentrations of high or low values (i.e., hot or cold spots) of social influence across the United States, for each election year. More specifically, the hotspot analysis of US presidential elections from 1920 to 2012 reveals a distinctive geographical cluster of states with statistically significant low levels of social influence (cold spots). This cluster is comprised of states mainly in the Great Plains and West North Central regions (including, for example, Montana, Wyoming, North Dakota, South Dakota, Nebraska, Kansas and Oklahoma). Contrastingly, states predominantly in the Middle Atlantic region (New Jersey, Pennsylvania, and New York)—for all election years—and states in the Pacific region (California, and Oregon) and the Southwest (Arizona and Nevada)—from 1988 to 2012—display high values of social influence (hot spots).

It would be interesting to speculate on the political, economic, social, and psychological factors that drive geographic variation in voting contagion. Research in the geographical and psychological sciences, which examine the geographical distribution of political, economic, social, and personality traits within the United States (67-72), suggest that the Great Plains and West North Central region is characterized by individuals that are typified by conservative social values, low openness and resistance to change, and preference of familiarity over novelty. This region comprises states with comparatively small minority populations (70), is less affluent, has fewer highly educated residents, is less innovative compared with other regions, and tends to be politically conservative and religious (69). Individual in this region choose to settle near family and friends and maintain intimate social relationships with them, but also tend to display low levels of social tolerance and acceptance for people who are from different cultures, unconventional, or live alternative lifestyles (69). Altogether, the above characteristics indicate a region where voters' choices are plausibly based upon strong ideology, party identification, orientations and attitudes rooted in religion and traditional social values, and reinforced by face-to-face interactions with like-minded family members and friends. We therefore expect our model to generate a social influence index (see Movie S1 for maps of raw social influence values instead of the Getis-Ord Local $G_i^*$ statistic) that reflects external forces (e.g., in the form of party identification or ideology), which are strong compared to peer influences.

Unlike the very low openness and conservative social values typical for the Great Plains and West North Central region, states along the Middle Atlantic and Southwest region are marked by moderately to very high openness, is wealthy, educated, culturally and ethnically diverse, and economically innovative (69, 71). This region appears to be politically liberal, and has fewer mainline Protestants (69). Residents of this region also appear to be tolerant and accepting of social and cultural differences (69). Considering the social diversity, tolerance, openness, and open-mindedness in this region, it is plausible that people's orientations and attitudes are influenced by the attitudes of others (69). This is consistent with our model, which shows high levels of social influence index (see maps of raw social influence values in Movie S1) that indicate peer influences that are strong compared to external forces in the form of attitudes or ideology. Although further research is needed to uncover the factors affecting social influence, it is plausible that economic, social, and psychological factors, as discussed above, can explain the geographical variability of social influence.

## Discussion

Many complex systems can be viewed as comprising of numerous interconnected units (e.g., voters) each of which responds independently to external forces (e.g., party identification), but is also affected by the states (e.g., opinions) of its connected units (14-15, 46-47). In such systems, the distribution of the states of the units (e.g., vote-share distribution) may change in characteristic ways depending on the strength of external influences relative to internal influences (14-15). Therefore, a key question is how to disentangle the effect of internal influences from that of exposure to external influences, given observational data about the phenomena we are trying to explain. This identification problem is important not only to the biological and physical sciences (e.g., ecosystems (73)), but also in the social sciences where the importance of social interactions in forming opinions and decisions has been emphasized (8, 31-32, 34, 74). The numerous opinions held by people can be formed either by information acquired by external sources or, when comprehensive search for information is costly, time-consuming, or difficult to acquire or process, formed by the attitudes of their peers in their own personal networks.

In this paper, we presented a general methodology for quantifying the degree of social and peer influence on the basis of given observational data. The methodology is based on an extended version of the voter model (14-15) that takes into account the effect of external forces, and is applied to a comprehensive data of US presidential elections from 1920 to 2012. An essential element in the model is social interaction between individual voters. The model includes two external influence parameters that reflect the bias in favor of one of two candidates. These free parameters are also interpreted as external factors that influence voting choices, such as attitudes and orientations, party identification, ideology, campaign persuasion, and exposure to television and newspapers. In addition to these external factors, voters are also influenced by the behavior of others.

Our model is validated in several ways. First, we derive the theoretical probability distribution of the vote-share per county, and find a remarkable fit between the theoretical result and the empirically observed county vote-share distributions. Our theoretical result is also consistent with observations in other countries (52). To our knowledge this is the first study that provides an analytical expression of the county vote-share distribution. Second, we examined the temporal dynamics of social influence by calculating the social influence index for each state and each election year. Our analysis reveals a distinct pattern of increasing social influence over 92 years (1920-2012) of US presidential elections. The 1984 election year represents the phase transition point from low (1920-1984) to high (1984-2012) levels of social contagion. The increasing levels of social influence at presidential elections suggest, in turn, the decline of bias induced by external forces (e.g., partisanship among voters), and an increasing of independence in voting behavior. Third, we examined how the geographic variation across states in social influence changes over time. This spatiotemporal analysis enables our model to reproduce two stable long-term periods of election years corresponding to two successive long-term periods of low and high levels of social contagion, in alignment with the 1984 phase transition finding. This suggests a new data-driven, large-scale systems approach of characterizing abrupt transitions of political events, which is based on critical realignment in the patterns of social contagion. Finally, we use the model to map the social contagion geography of the United States. Results from spatial analysis reveal robust differences among regions of the United States in terms of their social influence index. In particular, we identify two regions of 'hot' and 'cold' spots of social influence, each comprising states that are geographically close. We provided some evidence that statewide variation in social contagion may be linked to psychological, social, and economic factors.

More broadly the results suggest the growing role of social influence and contagion in shaping peoples' behaviors, tastes, and actions in a variety of real-life situations. Social influence and contagion will likely become increasingly evident as our society becomes more interconnected through the information superhighway and transport infrastructure networks. If we want to truly understand macro-level collective behavior in human systems—and perhaps devise ways by which human society can increase its collective wisdom—it will be important to develop practical and effective methods for measuring and monitoring the extent of social influence.

## Materials and Methods

### Dynamic network model of voting

Consider a network representing a county with $N$ nonpartisan voters (variable nodes) taking only the values of $0$ or $1$, representing support for candidate 0 or 1, respectively (e.g., Republican or Democrat). In addition, there are $N^0$ and $N^1$ partisan voters (frozen nodes) in state 0 and 1, respectively. At each time step, a variable node is selected at random; with probability $1-p$ the node copies the state of one of its connected neighbors, and with probability $p$ the state remains unchanged. The partisan nodes can also be interpreted as external perturbations, representing a variety of factors that influence voters' attitudes towards one of the two candidates (e.g., party identification and ideology). Analytically extending $N^0$ and $N^1$ to be real numbers enables modeling arbitrary strengths of external perturbations.

For a fully connected network the behavior of the system can be solved exactly as follows. The nodes are indistinguishable and the state of the network is fully specified by the number of nodes with internal state 1. Therefore, there are only $N+1$ distinguishable global states, which we denote $S_k, k = 0, 1, \cdots, N$. The state $S_k$ has $k$ variable nodes in state 1 and $N-k$ variable nodes in state 0. If $P_t(k)$ is the probability of finding the network in state $S_k$ at time $t$, then $P_{t+1}(k)$ can depend only on $P_t(k)$, $P_t(k+1)$ and $P_t(k-1)$. The probabilities $P_t(k)$ define a vector of $N+1$ components $\mathbf{P}_t$. The dynamics is described by the equation

$$P_{t+1}(k) = P_t(k)\left\{p + \frac{(1-p)}{N(N+N^0+N^1-1)}[k(k+N^1-1) + (N-k)(N+N^0-k-1)]\right\}$$

$$+ P_t(k-1)\frac{(1-p)}{N(N+N^0+N^1-1)}(k+N^1-1)(N-k+1)$$

$$+ P_t(k+1)\frac{(1-p)}{N(N+N^0+N^1-1)}(k+1)(N+N^0-k-1).$$

The term inside the first brackets gives the probability that the state $S_k$ does not change in that time step and is divided into two contributions: the probability $p$ that the node does not change plus the probability $1-p$ that the node does change but copies another node in the same state. In the latter case, the state of the node is 1 with probability $k/N$, and it may copy a different node in the same state with probability $(k - 1 + N^1)/(N + N^0 + N^1 - 1)$. Also, if the state of the selected node is 0, which has probability $(N - k)/N$, it may copy another node in state 0 with probability $(N - k - 1 + N^0)/(N + N^0 + N^1 - 1)$. The other terms are obtained similarly.

In terms of $\mathbf{P}_t$, the dynamics is described by the equation

$$\mathbf{P}_{t+1} = \mathbf{T}\mathbf{P}_t \equiv \left(1 - \frac{(1-p)}{N(N + N^0 + N^1 - 1)}\mathbf{A}\right)\mathbf{P}_t \qquad [8]$$

where the time evolution matrix $\mathbf{T}$, and also the auxiliary matrix $\mathbf{A}$, is tridiagonal. The non-zero elements of $\mathbf{A}$ are independent of $p$ and are given by

$$A_{k,k} = 2k(N - k) + N^1(N - k) + N^0 k$$

$$A_{k,k+1} = -(k + 1)(N + N^0 - k - 1) \qquad [9]$$

$$A_{k,k-1} = -(N - k + 1)(N^1 + k - 1)$$

The transition probability from state $S_M$ to $S_L$ after a time $t$ can be written as

$$P(L, t; M, 0) = \sum_{r=0}^{N} b_{rM} a_{rL} \lambda_r^t \qquad [10]$$

where $a_{rL}$ and $b_{rM}$ are the components of the right and left $r$-th eigenvectors of the evolution matrix, $\mathbf{a}_r$ and $\mathbf{b}_r$. Thus, the dynamical problem has been reduced to finding the right and left eigenvectors and eigenvalues of the time evolution matrix $\mathbf{T}$.

The eigenvalues $\lambda_r$ of $\mathbf{T}$ are given by

$$\lambda_r = 1 - \frac{(1-p)}{N(N + N^0 + N^1 - 1)} r(r - 1 + N^0 + N^1) \qquad [11]$$

and satisfy $0 \leq p \leq \lambda_r \leq 1$. The equation for $P(L, t; M, 0)$ shows that the asymptotic behavior of the network is determined only by the right and left eigenvectors with unit eigenvalue, i.e., by the eigenvector corresponding to $\lambda_0 = 1$. The coefficients of the corresponding (unnormalized) left eigenvector are simply $b_{0k} = 1$. The coefficients $a_{0k}$ of the right eigenvector are obtained using a generating function technique and an associated nonlinear second order differential equation (14-15). The coefficients are then given by the Taylor expansion of the hypergeometric function $F(-N, N^1, 1 - N - N^0, x) \equiv \sum_k a_{0k} x^k$. After normalization, these coefficients give the stationary distribution

$$\rho(k) = \frac{\binom{N^1+k-1}{k}\binom{N+N^0-k-1}{N-k}}{\binom{N+N^0+N^1-1}{N}}. \qquad [12]$$

This is the probability of finding the network with $k$ nodes in state 1 at equilibrium, and it is independent of the initial state. The other eigenvectors, corresponding to $\lambda_r \neq 1$, can also be calculated, and are also related to hypergeometric functions (14-15). Although these eigenvectors provide a complete description of the dynamics of the network (see Eq. 10), they are not particularly illuminating as we are interested in understanding the asymptotic behavior of the system ($\lambda_0 = 1$).

In the thermodynamic limit $N \to \infty$, we can define continuous variables $v = k/N$, $n^0 = N^0/N$ and $n^1 = N^1/N$ and approximate the asymptotic distribution presented in Eq. 12 by a Gaussian $\rho(v) = 1/\sqrt{2\pi\sigma^2} \rho_0 e^{[-(v-\mu)^2/2\sigma^2]}$ with mean $\mu = n^1/(n^0 + n^1) = N^1/(N^0 + N^1)$ and variance

$$\sigma^2 = \frac{n^0 n^1 (1 + n^0 + n^1)}{N(n^0 + n^1)^3}$$
$$= \mu(1-\mu)\left(\frac{1}{N^0 + N^1} + \frac{1}{N}\right) \qquad [13]$$

In the limit where $n^0, n^1 \gg 1$, the width depends only on the ratio $\alpha = n^0/n^1$ and is given by $\sqrt{\alpha/N}/(1 + \alpha)$. In particular, for $n^0 = n^1 \gg 1$, the width tends to $1/(2\sqrt{N})$.

While the model solved above was stated in terms of non-negative integer influence parameters $N^0, N^1$, it can be generalized to a model where the external influence parameters $N^0, N^1$ are real numbers. In this case, the solution in Eq. 12 remains the same, with the difference that factorials must be replaced by gamma functions. Since the numbers $N^0/(N + N^0 + N^1 - 1)$ and $N^1/(N + N^0 + N^1 - 1)$ represent the probabilities that a free node (nonpartisan voter) copies one of the frozen nodes (partisan voters), small (large) values of $N^0$ and $N^1$ can be interpreted as representing a weak (strong) connection between the free nodes and the external system containing the frozen nodes. The external system can be thought of as a reservoir that affects the network but is not affected by it.

**Model behavior**

Fig. 6 shows examples of the distribution $\rho(k)$ for a network with $N = 500$ and various values of $N^0$ and $N^1$. As we move around in the $(N^0, N^1)$-parameter space, we observe different types of behavior, which is characteristic of a first-order phase transition.

For $N^0 = N^1 = 1$, we obtain $\rho(k) = 1/(N+1)$ for all values of $N$, i.e. $N^0 = N^1 = 1$ is the *critical value* of this model. In this case, all states $S_k$ are equally likely and the system executes a random walk through the state space. In the limit $N \to \infty$, $N^0 = N^1 = 1$ marks the transition between disordered and ordered states.

For $N^0, N^1 > 1$, we obtain skewed unimodal distributions with peak at $N^1/(N^0 + N^1)$ corresponding to the fraction of voters in the network that voted for candidate 1. If $N^1 > N^0$, the majority of votes go to candidate 1, and if $N^0 > N^1$ the majority of votes go to candidate 0. We note that the estimation of the influence parameters $N^0, N^1$, based on almost a century of US presidential election data, predominantly falls within this regime. For $N^0, N^1 \gg 1$, $\rho(k)$ resembles a Gaussian distribution, and if $N^0 = N^1$ about half the voters vote for candidate 0 and half the voters vote for candidate 1, similarly to a magnetic material at high temperatures.

For $N^0, N^1 < 1$—the bistable (hysteresis) region—we obtain bimodal distributions in which either of the two network phases can exist, similar to the magnetization state in the Ising model below the critical temperature. For $N^0 = N^1 \ll 1$, the distribution peaks at all nodes 0 or all nodes 1, similar to a magnetized state at low temperatures.

Finally, for $N^1 > 1, N^0 < 1$ or $N^1 < 1, N^0 > 1$, we obtain unimodal distributions with peaks at all nodes 1 or all nodes 0, respectively.

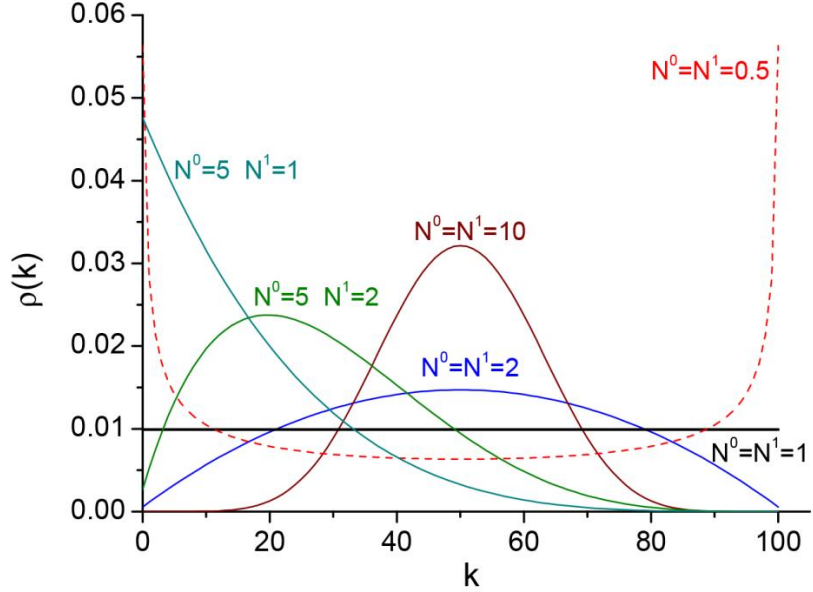

**Fig. 6. Stationary distributions for different values of $N^0$ and $N^1$.** Probability distributions of finding the network with $k$ nodes in state 1 at equilibrium for different values of $N^0$ and $N^1$. The number of variable nodes is $N = 500$.

**Other network topologies**

Although the stationary vote-share distribution given by Eq. 12 is obtained assuming fully connected networks, it was shown in (14-15) that our exact results are excellent approximations for other networks, including random, regular lattice, scale-free, and small world networks. These approximations can be useful, for example, if our model is applied to a network constructed based on online social networks or commuting networks. For these networks, which are not fully connected, the effect of the frozen nodes is amplified and can be quantified as follows: the probability that a free node copies a frozen node is $P_i = (N^0 + N^1)/(N^0 + N^1 + k_i)$ where $k_i$ is the degree of the node. We can then define effective numbers of frozen nodes in the corresponding fully connected network, $N^{0ef}$ and $N^{1ef}$, as being the values for which

$$\frac{(N^{0ef} + N^{1ef})}{(N^{0ef} + N^{1ef} + N - 1)} = \sum_k \frac{(N^0 + N^1)}{(N^0 + N^1 + k)} f(k) \quad [14]$$

$$\frac{N^{1ef}}{(N^{0ef} + N^{1ef})} = \frac{N^1}{(N^0 + N^1)} \quad [15]$$

where the term on the right-hand side in Eq. 14 is the expectation with respect to the degree distribution $f(k)$, and the term on the left-hand side is the probability that a free node copies a frozen node in the corresponding fully connected network. Eq. 15 is the mean field boundary condition. For well-behaved distributions, $N^{0ef}$ and $N^{1ef}$ can be obtained in terms of central moments of the degree distribution by expanding the right-hand side in Eq. 14 around the average degree $\langle k \rangle$ of the real network, as follows:

$$\frac{(N^{0ef} + N^{1ef})}{(N^{0ef} + N^{1ef} + N - 1)} = \sum_n (-1)^n \frac{(N^0 + N^1)}{(N^0 + N^1 + \langle k \rangle)^{(n+1)}} \mu_n \quad [16]$$

where $\mu_n = \sum (k - \langle k \rangle)^n f(k)$ are the central moments of the distribution $f(k)$. For example, using only the first term in the Taylor expansion gives $(N^{0ef} + N^{1ef})/(N^{0ef} + N^{1ef} + N - 1) = (N^0 + N^1)/(N^0 + N^1 + \langle k \rangle)$. This leads to

$$N^{0ef} = fN^0 \qquad N^{1ef} = fN^1$$

where $f = (N-1)/\langle k \rangle$. Therefore, as the network acquires more internal connections and $\langle k \rangle$ increases, the effective values $N^{0ef}$ and $N^{1ef}$ decrease.

## Supplementary Materials (downloaded at http://bit.ly/2dh0h12)

**Movie S1. Social influence topography of the United States: 1920-2012**. Movie S1 shows maps of social influence for all election years. The colored areas are derived from the social influence index calculated using Eq. 7 in the main text.

**Movie S2. Hotspots of social contagion: 90 years of presidential elections**. Movie S2 shows colored maps that reflect the significance (p-value) of local concentration of social influence for each state. The p-values for each state are derived from a random permutation test of local clustering using the Getis-Ord Local $G_i^*$ statistic. This analysis was performed with a contiguity spatial weight matrix that indicates whether states share a boundary or not. The variable of concern is the social influence index calculated using Eq. 7 in the main text. Low p-values (p-value≤0.1) indicate statistically significant high levels of social influence at a state and its surrounding neighbors (hot spots). High p-values (p-value≥0.9) indicate statistically significant low levels of social influence at a state and its surrounding neighbors (cold spots).

**Fig. S1. Testing for a break in the level of social influence using the Mann-Whiney U-test**.

**Fig. S2.** Hot spot analysis of social influence: 1920 US presidential election.

**Fig. S3.** Hot spot analysis of social influence: 1924 US presidential election.

**Fig. S4.** Hot spot analysis of social influence: 1928 US presidential election.

**Fig. S5.** Hot spot analysis of social influence: 1932 US presidential election.

**Fig. S6.** Hot spot analysis of social influence: 1936 US presidential election.

**Fig. S7.** Hot spot analysis of social influence: 1940 US presidential election.

**Fig. S8.** Hot spot analysis of social influence: 1944 US presidential election.

**Fig. S9.** Hot spot analysis of social influence: 1948 US presidential election.

**Fig. S10.** Hot spot analysis of social influence: 1952 US presidential election.

**Fig. S11.** Hot spot analysis of social influence: 1956 US presidential election.

**Fig. S12.** Hot spot analysis of social influence: 1960 US presidential election.

**Fig. S13.** Hot spot analysis of social influence: 1964 US presidential election.

**Fig. S14.** Hot spot analysis of social influence: 1968 US presidential election.

**Fig. S15.** Hot spot analysis of social influence: 1972 US presidential election.

**Fig. S16.** Hot spot analysis of social influence: 1976 US presidential election.

**Fig. S17.** Hot spot analysis of social influence: 1980 US presidential election.

**Fig. S18.** Hot spot analysis of social influence: 1984 US presidential election.

**Fig. S19.** Hot spot analysis of social influence: 1988 US presidential election.

**Fig. S20.** Hot spot analysis of social influence: 1992 US presidential election.

**Fig. S21.** Hot spot analysis of social influence: 1996 US presidential election.

**Fig. S22.** Hot spot analysis of social influence: 2000 US presidential election.

**Fig. S23.** Hot spot analysis of social influence: 2004 US presidential election.

**Fig. S24.** Hot spot analysis of social influence: 2008 US presidential election.

**Fig. S25.** Hot spot analysis of social influence: 2012 US presidential election.

**Table S1.** Results of random permutation tests of spatial autocorrelation using Moran's $I$ statistic.

**Table S2.** Results of random permutation tests of spatial clustering using Getis-Ord General G* statistic.